\documentclass{quantumview}
\usepackage[utf8]{inputenc}
\usepackage[english]{babel}
\usepackage[T1]{fontenc}
\usepackage{amsmath}
\usepackage{hyperref}
\usepackage{fixltx2e}
\usepackage[numbers,sort&compress]{natbib}

\begin{document}

\title{Stretching the limits of multiparticle entanglement}

\author{G\'eza T\'oth}
\affiliation{Department of Theoretical Physics, University of the Basque Country UPV/EHU, \\ P. O. Box 644, E-48080 Bilbao, Spain}
\affiliation{IKERBASQUE, Basque Foundation for Science, E-48013 Bilbao, Spain}
\affiliation{Wigner Research Centre for Physics, P. O. Box 49, H-1525 Budapest, Hungary} 
\orcid{0000-0002-9602-751X}

\newcommand{\bra}[1]{\langle #1|}
\newcommand{\ket}[1]{|#1\rangle}
\newcommand{\ketbra}[1]{| #1\rangle \langle #1|}
\newcommand{\C}{\ensuremath{\mathbbm C}}
\newcommand{\va}[1]{\ensuremath{(\Delta#1)^2}}
\newcommand{\vasq}[1]{\ensuremath{[\Delta#1]^2}}
\newcommand{\varho}[1]{\ensuremath{(\Delta_\rho #1)^2}}
\newcommand{\ex}[1]{\ensuremath{\langle{#1}\rangle}}
\newcommand{\exs}[1]{\ensuremath{\langle{#1}\rangle}}
\newcommand{\mean}[1]{\ensuremath{\langle{#1}\rangle}}
\newcommand{\exrho}[1]{\ensuremath{\left\langle{#1}\right\rangle}_{\rho}}
\newcommand{\eins}{\mathbbm{1}}
\newcommand{\WW}{\ensuremath{\mathcal{W}}}
\newcommand{\NN}{\ensuremath{\mathcal{N}}}
\newcommand{\XX}{\ensuremath{\mathcal{X}}}
\newcommand{\MM}{\ensuremath{\mathcal{M}}}
\newcommand{\HH}{\ensuremath{\mathcal{H}}}
\newcommand{\FF}{\ensuremath{\mathcal{F}}}
\newcommand{\PP}{\ensuremath{\mathcal{P}}}
\renewcommand{\AA}{\ensuremath{\mathcal{A}}}
\newcommand{\BB}{\ensuremath{\mathcal{B}}}
\newcommand{\QQ}{\ensuremath{\mathcal{Q}}}
\newcommand{\VV}{\ensuremath{\mathcal{V}}}
\newcommand{\LL}{\ensuremath{\mathcal{L}}}
\newcommand{\SSS}{\ensuremath{\mathcal{S}}}
\newcommand{\EE}{\ensuremath{\mathcal{E}}}
\newcommand{\OO}{\ensuremath{\mathcal{O}}}
\newcommand{\fO}{\ensuremath{\mathfrak{O}}}
\newcommand{\kommentar}[1]{}
\newcommand{\trace}{{\rm Tr}}
\newcommand{\dt}{\ensuremath{\partial_t}}
\renewcommand{\aa}{\ensuremath{\alpha}}
\newcommand{\bb}{\ensuremath{\beta}}
\newcommand{\vr}{\ensuremath{\varrho}}
\newcommand{\forget}[1]{}
\newcommand{\tr}{\mbox{Tr}}
\newcommand{\EQ}[1]{Eq.~\eqref{#1}}
\newcommand{\EQS}[1]{Eqs.~\eqref{#1}}
\newcommand{\EQL}[1]{Equation~\eqref{#1}}
\newcommand{\SEC}[1]{Sec.~\ref{#1}}
\newcommand{\FIG}[1]{Fig.~\ref{#1}}
\newcommand{\REF}[1]{Ref.~\cite{#1}}
\newcommand{\REFS}[1]{Refs.~\cite{#1}}
\newcommand{\REFL}[1]{Reference~\cite{#1}}
\newcommand{\REFSL}[1]{References~\cite{#1}}
\newcommand{\APP}[1]{Appendix~\ref{#1}}
\newcommand{\lA}{\mathrm{A}}
\newcommand{\lB}{\mathrm{B}}
\newcommand{\lC}{\mathrm{C}}
\newcommand{\lD}{\mathrm{D}}

\maketitle

The classification of entangled mixed states in multiparticle systems is a difficult task. Even for three-qubit pure states, six classes arise that are inequivalent under Stochastic Local Operations and Classical Communication (SLOCC), which can be used to define a classification of mixed states into six classes \cite{Acin2001Classification}. For pure states of more than three particles, there are already infinite number of classes \cite{Verstraete2002Four}. So it is desirable to find coarser classifications.

One of the possible classifications is the following. We look for the largest particle group that contains particles entangled with each other while it is non-entangled with the rest. Then, we call the state $5$, $10$ or $50$ particle entangled based on this \cite{Sorensen2001Entanglement,Guhne2005Multipartite}. The entanglement depth or $k$-producibility has been defined this way. In more detail, we call a pure state of $N$ particles $k$-producible, if it can be written as 
\begin{equation}
\vert \Psi_{1}\rangle \otimes \vert \Psi_{2}\rangle \otimes \vert \Psi_{3}\rangle \otimes \dots,
\label{eq:product}
\end{equation}
where each $\ket{\Psi_{l}}$ is of the state of at most $k$ particles. A mixed state is $k$-producible if it is the mixture of pure $k$-producible states. If a quantum state is not $k$-producible  then it is at least $(k+1)$-particle entangled, or it has at least an entanglement depth $k+1$.

There have been many groundbreaking experiments putting a lower bound on the entanglement depth of the quantum system, aiming to produce larger and larger entanglement depth, creating entanglement depth in the thousands \cite{Gross2010Nonlinear,Lucke2014Detecting,Hosten2016Measurement,McConnell2015Entanglement,Haas2014Entangled,Zou2018Beating}. At this point, an important question arises. If we have 100 particles in a $20$-particle entangled state, it can happen in various ways. For example, it can happen, that all twenty-particle groups are fully entangled
\begin{equation}
\Bigl[\frac1 {\sqrt 2}\bigl(\vert0\rangle^{\otimes 20}+\ket{1}^{\otimes 20}\bigr)\Bigr] ^{\otimes 5},
\label{eq:state1}
\end{equation}
or it can also happen that there is a single twenty-particle group that is genuine multiparticle entangled, while the rest of the particles are in the trivial $\vert0\rangle$ state
\begin{equation}
\Bigl[\frac1 {\sqrt 2}\bigl(\vert0\rangle^{\otimes 20}+\ket{1}^{\otimes 20}\bigr)\Bigr]  \otimes \vert0\rangle^{\otimes 80}.
\label{eq:state2}
\end{equation}
It is natural to ask what further notions in entanglement theory can be used to distinguish these two cases.

The article of Sz.~Szalay from the Wigner Research Centre for Physics in Budapest published in Quantum \cite{Szalay2019k} is just doing that. The preliminary ideas, laid down in previous works \cite{Szalay2012MultipartEntClass,Szalay2015MultipartEntMeasures,Szalay2017ChemBonds}, are as follows. First, on level I, the author characterizes the total system by the use of its \emph{partitions,}
\begin{equation}
\xi=X_1|X_2|X_3|\dots,
\label{eq:part}
\end{equation}
where a part $X_l$ is a subsystem, possibly consisting of several elementary subsystems (e.g., particles). \emph{$\xi$-uncorrelated states} are just product states of the form
\begin{equation}
\varrho_{1} \otimes \varrho_{2} \otimes \varrho_{3} \otimes \dots,
\label{eq:product3}
\end{equation}
where $\varrho_l$ lives on subsystem $X_l$. \emph{$\xi$-separable states} are those which can be formed as mixtures of $\xi$-uncorrelated states, that is, states that are separable for the partitioning given by $\xi$. After the level I, the article defines level II and level III descriptions  using fundamental set theory that can handle in a coherent way a large variety of relevant cases appearing in multiparticle systems. Level II is needed for handling mixtures of states uncorrelated with respect to different partitions. For example, considering three elementary subsystems $\lA$, $\lB$ and $\lC$, the $\{\lA\lB|\lC,\lB\lC|\lA,\lA\lC|\lB\}$-separable states are mixtures of  $\lA\lB|\lC$-uncorrelated, $\lB\lC|\lA$-uncorrelated and $\lA\lC|\lB$-uncorrelated states. These states are not considered tripartite entangled \cite{Acin2001Classification,Seevinck2008partsep}.

So level II is about the possible states from which the state can be mixed, then level III is about the possible states from which the state can be mixed \emph{and} from which it cannot be mixed. For examples, besides the cases known earlier \cite{Acin2001Classification,Seevinck2008partsep}, we mention that there are states that are $\{\lA\lB|\lC,\lB\lC|\lA,\lA\lC|\lB\}$-separable, but neither $\{\lA\lB|\lC,\lB\lC|\lA\}$-separable, nor $\{\lA\lB|\lC,\lA\lC|\lB\}$-separable, nor $\{\lB\lC|\lA,\lA\lC|\lB\}$-separable; that is, to mix them, shared bipartite entanglement is needed in all the three bipartite subsystems \cite{Szalay2012MultipartEntClass,Szalay2013dissertation}. Another example is that of the states which are $\lA\lB|\lC$-separable (while not being $\lA|\lB|\lC$-separable) and also $\{\lB\lC|\lA,\lA\lC|\lB\}$-separable; that is, they can be mixed without shared bipartite entanglement in $\lA\lB$, if we have shared bipartite entanglement in $\lB\lC$ and in $\lA\lC$. Such ``roundabout'' states \cite{Szalay2012MultipartEntClass,Szalay2013dissertation} were constructed only recently \cite{Han2018MultipartEntClasses3qb}.

In a nutshell, levels I and II describe the possible multipartite correlation and entanglement \emph{properties}, and level III is about the \emph{classification} in the strict sense. For the multipartite properties, correlation and entanglement measures are also constructed, generalizing the \emph{mutual information} and the \emph{entanglement of formation} or \emph{relative entropy of entanglement} for the mutipartite scenario. Partial orders can be defined for the different sets on all the three levels,  which give the structure of the notions on the different levels, and can be expressed in diagrams. The interesting properties of the structure of these is also the aim of recent research \cite{Han2019NonDistrib}. This approach is compatible with the LO paradigm for correlation and the LOCC paradigm entanglement. The state sets arising on levels I and II are closed with respect to LO/LOCC, the measures are correlation/entanglement monotones, and on level III the partial order goes along the LO/LOCC convertibility among the classes.

The novel results of the manuscript, concerning the case when only \emph{permutation invariant properties} are taken into account, fit well into this general framework. This restriction is well motivated when an ensemble of particles is described, which cannot be addressed one by one. For this, the same construction as the above is built up, but now based on \emph{integer partitions},
\begin{equation}
\hat{\xi}=x_1|x_2|x_3|\dots,
\label{eq:intpart}
\end{equation}
where $x_l$ is a possible subsystem-size (e.g., particle number).
\emph{$\hat{\xi}$-uncorrelated states} are just product states of the form
\begin{equation}
\varrho_{1} \otimes \varrho_{2} \otimes \varrho_{3} \otimes \dots,
\label{eq:product2}
\end{equation}
where now $\varrho_l$ lives on a subsystem of size $x_l$, not specified, which one. \emph{$\hat{\xi}$-separable states} are those which can be formed as mixtures of $\hat{\xi}$-uncorrelated states. Again, level II and level III descriptions can be formulated analogously.

Here comes an elaborate definition of $k$-producibility and $k$-partitionability in the above picture. A partition of the type \eqref{eq:part} is \emph{$k$-producible}, if all the subsystems contain \emph{at most} $k$ elementary subsystems, e.g., particles. A partition is \emph{$k$-partitionable}, if the number of subsystems is \emph{at least} $k$. Then, we can talk about $k$-producibly uncorrelated and $k$-producibly separable states. We can also define $k$-partitionably uncorrelated and $k$-partitionably separable states. (The $k$-producibly separable states are called $k$-producible states in entanglement theory, while $k$-partitionibly separable states are $k$-separable states. The article uses the more general naming, because it considers correlation and entanglement in parallel, and the name ``$k$-separably uncorrelated'' would not make sense.)
 
Then, \emph{Young diagrams} are used to represent the permutationally invariant case. This is very expressive: what matters is to know how many times the various subsystem sizes appear, and it is not important, which elementary subsystems a given subsystem consists of. In a Young diagram, every row of $x_l$ squares indicates a group of $x_l$ elementary subsystems forming a subsystem. A Young diagram of horizontal size $k$ and vertical size $k'$ correspond to a partition being $k$-producible and $k'$-partitionable. The \emph{conjugation} of Young diagrams, which is the flip with respect to the diagonal, interchanges the horizontal and vertical sizes, establishing an interesting \emph{duality}, connecting producibility and partitionability. 

Finally, \emph{stretchability}, appearing in the title, is the difference of producibility and partitionability. Hence, the Young diagram mentioned above would have a stretchability $k-k'$. The notions can be clearly understood based on Figure 6 in \cite{Szalay2019k}. All these can be applied to define the stretchability for correlation and for entanglement, combining the advantages of producibility and partitionability in a balanced way. For $N$ particles, the stretchability of entanglement $N-1$, if the state is fully $N$-partite entangled.   The stretchability of entanglement is $-(N-1)$, if the state is fully separable. $k$-stretchability combines the advantages of $k$-partitionability and $k$-producibility, it is large if there are a small number of large correlated or entangled subsystems, and it is small, if the subsystems are smaller, or if there are too many of them. For example, for the states \eqref{eq:state1} and \eqref{eq:state2}, the stretchability is $+15$ and $-61$, respectively. Also, stretchability of correlation/entanglement is decreasing for LO/LOCC. In short, $k$-stretchability is defined as a new quantity added to $k$-producibility and $k$-separability to characterize better the multipartite entanglement of the quantum state.

\bibliography{QuantumViews19_v11}

\begin{thebibliography}{10}

\bibitem{Acin2001Classification}
A.~Ac{\'{\i}}n, D.~Bru\ss{}, M.~Lewenstein, and A.~Sanpera.
\newblock ``Classification of mixed three-qubit states''.
\newblock \href{https://dx.doi.org/10.1103/PhysRevLett.87.040401}{Phys. Rev.
  Lett. {\bf 87}, 040401}~(2001).

\bibitem{Verstraete2002Four}
F.~Verstraete, J.~Dehaene, B.~De~Moor, and H.~Verschelde.
\newblock ``Four qubits can be entangled in nine different ways''.
\newblock \href{https://dx.doi.org/10.1103/PhysRevA.65.052112}{Phys. Rev. A
  {\bf 65}, 052112}~(2002).

\bibitem{Sorensen2001Entanglement}
Anders~S. S\o{}rensen and Klaus M\o{}lmer.
\newblock ``Entanglement and extreme spin squeezing''.
\newblock \href{https://dx.doi.org/10.1103/PhysRevLett.86.4431}{Phys. Rev.
  Lett. {\bf 86}, 4431--4434}~(2001).

\bibitem{Guhne2005Multipartite}
Otfried G{\"u}hne, G\'eza T\'oth, and Hans~J Briegel.
\newblock ``Multipartite entanglement in spin chains''.
\newblock \href{https://dx.doi.org/10.1088/1367-2630/7/1/229}{New J. Phys. {\bf
  7}, 229}~(2005).

\bibitem{Gross2010Nonlinear}
Christian Gross, Tilman Zibold, Eike Nicklas, Jerome Esteve, and Markus~K
  Oberthaler.
\newblock ``Nonlinear atom interferometer surpasses classical precision
  limit''.
\newblock \href{https://dx.doi.org/10.1038/nature08919}{Nature (London) {\bf
  464}, 1165--1169}~(2010).

\bibitem{Lucke2014Detecting}
Bernd L\"ucke, Jan Peise, Giuseppe Vitagliano, Jan Arlt, Luis Santos, G\'eza
  T\'oth, and Carsten Klempt.
\newblock ``Detecting multiparticle entanglement of {D}icke states''.
\newblock \href{https://dx.doi.org/10.1103/PhysRevLett.112.155304}{Phys. Rev.
  Lett. {\bf 112}, 155304}~(2014).

\bibitem{Hosten2016Measurement}
O.~{Hosten}, N.~J. {Engelsen}, R.~{Krishnakumar}, and M.~A. {Kasevich}.
\newblock ``{Measurement noise 100 times lower than the quantum-projection
  limit using entangled atoms}''.
\newblock \href{https://dx.doi.org/10.1038/nature16176}{Nature (London) {\bf
  529}, 505--508}~(2016).

\bibitem{McConnell2015Entanglement}
R.~{McConnell}, H.~{Zhang}, J.~{Hu}, S.~{{\'C}uk}, and V.~{Vuleti{\'c}}.
\newblock ``{Entanglement with negative Wigner function of almost 3,000 atoms
  heralded by one photon}''.
\newblock \href{https://dx.doi.org/10.1038/nature14293}{Nature (London) {\bf
  519}, 439--442}~(2015).

\bibitem{Haas2014Entangled}
Florian Haas, J\"urgen Volz, Roger Gehr, Jakob Reichel, and Jerome Esteve.
\newblock ``Entangled states of more than 40 atoms in an optical fiber
  cavity''.
\newblock \href{https://dx.doi.org/10.1126/science.1248905}{Science {\bf 344},
  180--183}~(2014).

\bibitem{Zou2018Beating}
Yi-Quan Zou, Ling-Na Wu, Qi~Liu, Xin-Yu Luo, Shuai-Feng Guo, Jia-Hao Cao,
  Meng~Khoon Tey, and Li~You.
\newblock ``Beating the classical precision limit with spin-1 {D}icke states of
  more than 10,000 atoms''.
\newblock \href{https://dx.doi.org/10.1073/pnas.1715105115}{Proc. Natl. Acad.
  Sci. U.S.A. {\bf 115}, 6381--6385}~(2018).

\bibitem{Szalay2019k}
{\relax Sz}il{\'a}rd {\relax Sz}alay.
\newblock ``$k$-stretchability of entanglement, and the duality of
  $k$-separability and $k$-producibility''.
\newblock \href{https://dx.doi.org/10.22331/q-2019-12-02-204}{Quantum {\bf 3},
  204}~(2019).

\bibitem{Szalay2012MultipartEntClass}
{\relax Sz}il{\'a}rd {\relax Sz}alay and Zolt{\'a}n K{\"o}k{\'e}nyesi.
\newblock ``Partial separability revisited: Necessary and sufficient
  criteria''.
\newblock \href{https://dx.doi.org/10.1103/PhysRevA.86.032341}{Phys. Rev. A
  {\bf 86}, 032341}~(2012).

\bibitem{Szalay2015MultipartEntMeasures}
{\relax Sz}il{\'a}rd {\relax Sz}alay.
\newblock ``Multipartite entanglement measures''.
\newblock \href{https://dx.doi.org/10.1103/PhysRevA.92.042329}{Phys. Rev. A
  {\bf 92}, 042329}~(2015).

\bibitem{Szalay2017ChemBonds}
{\relax Sz}il{\'a}rd {\relax Sz}alay, Gergely Barcza, Tibor {\relax
  Sz}ilv{\'a}si, Libor Veis, and {\"O}rs Legeza.
\newblock ``The correlation theory of the chemical bond''.
\newblock \href{https://dx.doi.org/10.1038/s41598-017-02447-z}{Sci. Rep. {\bf
  7}, 2237}~(2017).

\bibitem{Seevinck2008partsep}
Michael Seevinck and Jos Uffink.
\newblock ``Partial separability and entanglement criteria for multiqubit
  quantum states''.
\newblock \href{https://dx.doi.org/10.1103/PhysRevA.78.032101}{Phys. Rev. A
  {\bf 78}, 032101}~(2008).

\bibitem{Szalay2013dissertation}
{\relax Sz}il{\'a}rd {\relax Sz}alay.
\newblock ``Quantum entanglement in finite-dimensional {H}ilbert
  spaces''~(2013).
\newblock  \href{http://arxiv.org/abs/1302.4654}{arXiv:1302.4654}.

\bibitem{Han2018MultipartEntClasses3qb}
Kyung~Hoon Han and Seung-Hyeok Kye.
\newblock ``Construction of three-qubit biseparable states distinguishing kinds
  of entanglement in a partial separability classification''.
\newblock \href{https://dx.doi.org/10.1103/PhysRevA.99.032304}{Phys. Rev. A
  {\bf 99}, 032304}~(2019).

\bibitem{Han2019NonDistrib}
Kyung~Hoon Han, Seung-Hyeok Kye, and {\relax Sz}il{\'a}rd {\relax Sz}alay.
\newblock ``Partial separability/entanglement violates distributive
  rules''~(2019).
\newblock  \href{http://arxiv.org/abs/1911.06496}{arXiv:1911.06496}.

\end{thebibliography}

\bibliographystyle{quantum}

\end{document}